\documentclass{aa}  
\usepackage[breaklinks,colorlinks,citecolor=blue]{hyperref}
\usepackage{graphicx}
\usepackage{txfonts}
\usepackage{color}
\usepackage[flushleft]{threeparttable}
\begin{document} 

\newcommand\blfootnote[1]{%
  \begingroup
  \renewcommand\thefootnote{}\footnote{#1}%
  \addtocounter{footnote}{-1}%
  \endgroup
}

   \title{Characterization of foreground emission at degree angular scale for CMB B-modes observations
}

   \subtitle{Thermal Dust and Synchrotron signal from Planck and WMAP data}

   \author{N. Krachmalnicoff\inst{1,2}
          \and
          C. Baccigalupi\inst{3,4}
           \and
          J. Aumont\inst{5}
          \and
          M. Bersanelli\inst{1,2}
          \and
          A. Mennella\inst{1,2}
          }

   \institute{Dipartimento di Fisica, Universit\`a degli Studi di Milano, Via
Celoria, 16, Milano, Italy
         \and
         INAF/IASF Milano, Via E. Bassini 15, Milano, Italy
         \and
             SISSA, Astrophysics Sector, via Bonomea 265, 34136, Trieste, Italy
             \and
             INFN, Via Valerio 2, 34127, Trieste, Italy
             \and
             Institut d'Astrophysique Spatiale, CNRS (UMR8617) Universit\' e
Paris-Sud 11, B\^ atiment 121, Orsay, France
                          }

   \date{Preprint online version: November 2, 2015}

 
 \abstract
{We quantify the contamination from polarized diffuse Galactic synchrotron and thermal dust emissions to   
the $B$-modes of the Cosmic Microwave Background (CMB) anisotropies on the degree angular scale, using data from the \textit{Planck} and Wilkinson Microwave Anisotropy Probe (WMAP) satellites. We compute power spectra of foreground polarized emissions in 352 circular sky patches located at Galactic latitude $|b|>20^{\circ}$, each of which covering a fraction of the sky of about 1.5\%.
We make use of the  spectral properties derived from \textit{Planck} and WMAP data to extrapolate, in frequency, the amplitude of synchrotron and thermal dust $B$-modes spectra in the multipole bin centered at $\ell\simeq80$. In this way we estimate, for each analyzed region, the amplitude and frequency of the foreground minimum. We detect both dust and synchrotron signal, at degree angular scale and at $3\sigma$ confidence level, in 28 regions. 
Here the minimum of the foreground emission is found at frequencies between 60 and 100 GHz with an amplitude, expressed in terms of the equivalent tensor-to-scalar ratio, $r_{FG, \text{min}}$, between $\sim0.06$ and $\sim1$. Some of these regions are located at high Galactic latitudes, in areas close to the ones which are being observed by sub-orbital experiments. 
In all the other sky patches, where synchrotron or dust $B$-modes are not detectable with the required confidence, we put upper limits on the minimum foreground contamination and find values of $r_{FG, \text{min}}$ between $\sim0.05$ and $\sim1.5$, in the frequency range 60-90 GHz. Our results indicate that, with the current sensitivity at low frequency, it is not possible to exclude the presence of synchrotron contamination to CMB cosmological $B$-modes at the level requested to measure a gravitational waves signal with $r\simeq0.01$, at frequency $\lesssim100$ GHz, anywhere. 
Therefore, more accurate data are essential in order to better characterize the synchrotron polarized component, and eventually, remove its contamination to CMB signal through foreground cleaning.}

   {}
   {}
   {}
   {}
   {}


   \titlerunning{Characterization of foreground emission for CMB B-modes observations}
   \maketitle
%

\section{Introduction}
 \blfootnote{\textit{Send offprint requests to}: Nicoletta Krachmalnicoff, \href{mailto:nicoletta.krachmalnicoff@unimi.it}{nicoletta.krachmalnicoff@unimi.it}}
The study and characterization of the polarized Cosmic Microwave Background (CMB) signal represent one of the greatest challenges for modern observational cosmology. In particular, the existence of tensor perturbations in the CMB temperature and polarization signal, generated by primordial gravitational waves (GWs), is predicted within the 
inflationary scenario of the Early Universe, where a transient vacuum energy phase provides accelerated expansion and quantum generation of cosmological perturbations. 
The amplitude of these perturbations is often parametrized by the tensor-to-scalar ratio $r$. Measurements, or constraints, of this quantity are essential to discriminate among different theories 
\citep{planck2015-XX}. \par 
In recent years, several experiments have been designed and built with the major scientific goal of detecting the effect of GWs into CMB anisotropies, focusing, in particular, 
on the characterization of the curl component of the CMB polarized signal, the $B$-mode (see \citealp{B2P} and references therein). The contribution from primordial GWs is relevant
at angular scales larger than about one degree. On arcminute scales, the dominant contribution is given by Gravitational Lensing (GL) due to deflection of CMB photons 
by cosmological structures along their travel to the observer. 
Experiments looking at CMB $B$-modes include the European Space Agency (ESA) \textit{Planck} satellite \citep{planck2015-I}, ground-based experiments, 
like PolarBear \citep{PolarBear14}, BICEP2 \citep{BICEP2}, the Keck Array \citep{B2K}, QUIET \citep{QUIET}, QUBIC \citep{QUBIC} and CLASS \citep{CLASS}, 
and balloon-borne instruments, like EBEX \citep{EBEX}, SPIDER \citep{SPIDER}  and LSPE \citep{LSPE12})\footnote{A complete list of operating and planned 
CMB experiments is available at \url{ http://lambda.gsfc.nasa.gov/product/expt/}}. \par
These experimental efforts have provided us with upper limits on the value of $r$.  The most stringent ones come from \textit{Planck} temperature and 
polarization data, yielding to $r<0.11$ at 95\% C.L. \citep{planck2015-XIII}, and from the BICEP2/Keck/\textit{Planck} joint analysis of the $B$-mode polarized signal, 
resulting in $r<0.12$ at 95\% C.L. \citep{B2P}. The latter result is particularly relevant, as it comes from direct $B$-modes measurements. While this work was completed, a slightly more stringent upper limit was published, exploiting the latest data from the Keck array and BICEP2 telescopes, with $r<0.09$ at 95\% C.L. \citep{B2K2015}  \par
The detection of the CMB cosmological $B$-mode signal, at the angular scales where it should dominate over the signature generated by GL, is made difficult not only by 
the faintness of the signal itself (fraction of $\mu$K), but also by the presence of Galactic polarized foregrounds. \par
Two main kinds of Galactic foreground radiation emit strong linearly polarized signal: the thermal emission of Galactic dust and the synchrotron radiation. For a detailed description of these 
processes we refer to \citet{planck2015-X}, \citet{planck2014-XXX} and \citet{planck2015-XXV}.\par
Thermal radiation from the Galactic interstellar medium dominates the sky signal at frequencies above about 100 GHz and it is partially linearly polarized. The polarization fraction increases with  Galactic latitude and can reach the  $\sim$ 20\% level 
in several large regions of the sky (\citealp{planck2014-XIX}). Its frequency spectrum of thermal dust radiation is well described by a grey body,  $I_d(\nu)\propto\nu^{\beta_d}B_{\nu}(T_d)$, where $B_{\nu}(T_d)$ is the black body spectrum at 
temperature $T_{d}$. 
In \citet{planck2014-XXII} the \textit{Planck} and WMAP data are used to determine the spectral index $\beta_d$ of the thermal dust 
emission both in intensity and polarization, finding $\beta_d \simeq1.59$ for $T_d\simeq19.6$ K.\par
Synchrotron emission results from the acceleration of cosmic-ray electrons in the Galactic magnetic field. It dominates the sky signal at low frequency ($\lesssim$100 GHz). 
It is strongly linearly polarized, up to $\sim$20\% at intermediate and high Galactic latitudes (\citealp{Kogut07}), and shows a power-law frequency dependence 
with mean spectral index $\beta_s\simeq-3$ \citep{Fuskeland14}.\par
The relevance of foreground emission as contaminant for CMB polarization measurements, potentially at all frequencies and all Galactic latitudes, has been claimed in early studies \citep{Baccigalupi03} 
and confirmed, on large sky fractions, by WMAP observations  \citep{Page07, Gold11} and by \textit{Planck} data \citep{planck2015-X}. Recently \citet{Choi15} have used \textit{Planck} and WMAP data to study the spatial correlation between synchrotron and dust emission, finding a positive correlation with $\rho\approx0.4$ at large angular scales ($\ell<20$ and $f_{\text{sky}}>0.5$). They also found that, considering large portions of the sky, the minimum of foreground emission lies at frequency around 75 GHz and stressed how new data of synchrotron radiation, with higher sensitivity, are needed in order to completely understand the actual foreground contamination at frequency around 90 GHz.
The analysis reported in \citet{B2P} indicates that most, if not all, the $B$-mode excess measured by the BICEP2/Keck array data, 
originally interpreted as GWs signature \citep{BICEP2}, has to be assigned to the Galactic foreground emission. This result confirms, once again, how the characterization of polarized foregrounds represents a crucial aspect 
for current and future observation of the polarized CMB signal. \par
In \citet{planck2014-XXX} (hereafter PIP-XXX) the characterization of the polarized thermal dust radiation at intermediate and high Galactic latitudes is reported, focusing, in particular, 
on its relevance for CMB polarization measurements. In this work we extend that analysis also including  synchrotron radiation, largely based 
on the data taken in the 30 GHz band by the Low Frequency Instrument (LFI) onboard \textit{Planck}, and the WMAP-$K$ band at 22 GHz. Our goal is to describe the contamination from foreground emissions at degree angular scales exploiting the current available data. This, in particular, would be useful to optimize 
the observation strategies (e.g. frequency channels and observing region) for forthcoming and future sub-orbitals CMB experiments. To achieve this goal we analyze thermal dust and synchrotron polarization power 
spectra, with a common methodology, on 352 sky regions located at intermediate and high Galactic latitudes. In our analysis we  reproduce the PIP-XXX results on thermal dust radiation and we combine them with information about synchrotron radiation. In this way we estimate, for each considered sky region, the expected level and range of frequencies where the foreground 
emission reaches its minimum.  \par
The paper is organized as follows. In Section \ref{Sec_data} we present the data used for the analysis and we briefly describe also the {\tt Xpure} power spectrum estimator. In Section \ref{Sec:FG_cont} we report our analysis procedure and discuss the results concerning the impact of foreground emission on CMB 
$B$-modes measurements. Section \ref{Sec_conc} summarizes the main conclusions.

 \begin{figure}
   \centering
   \includegraphics[width=8.5 cm]{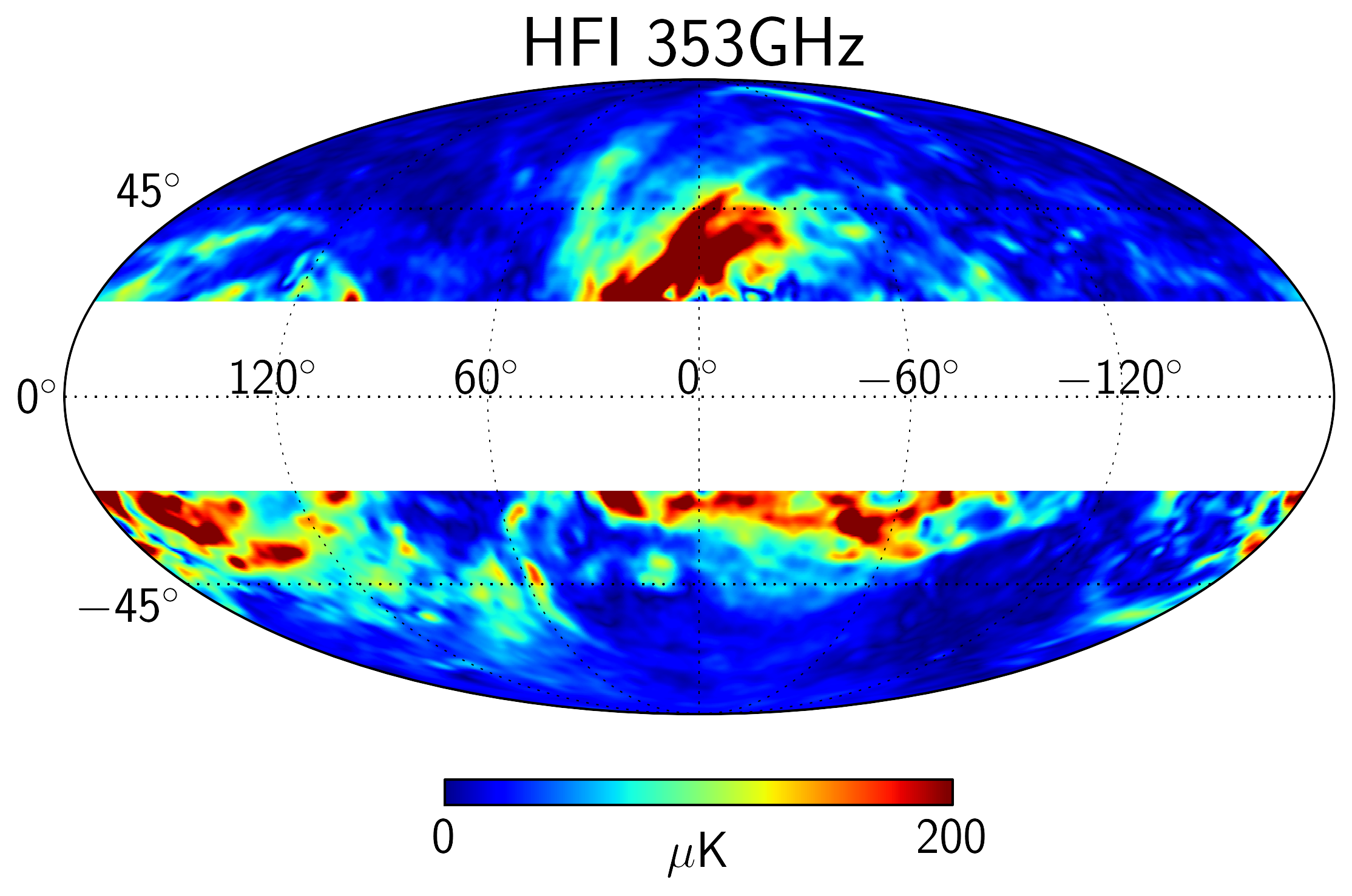}
    \includegraphics[width=8.5 cm]{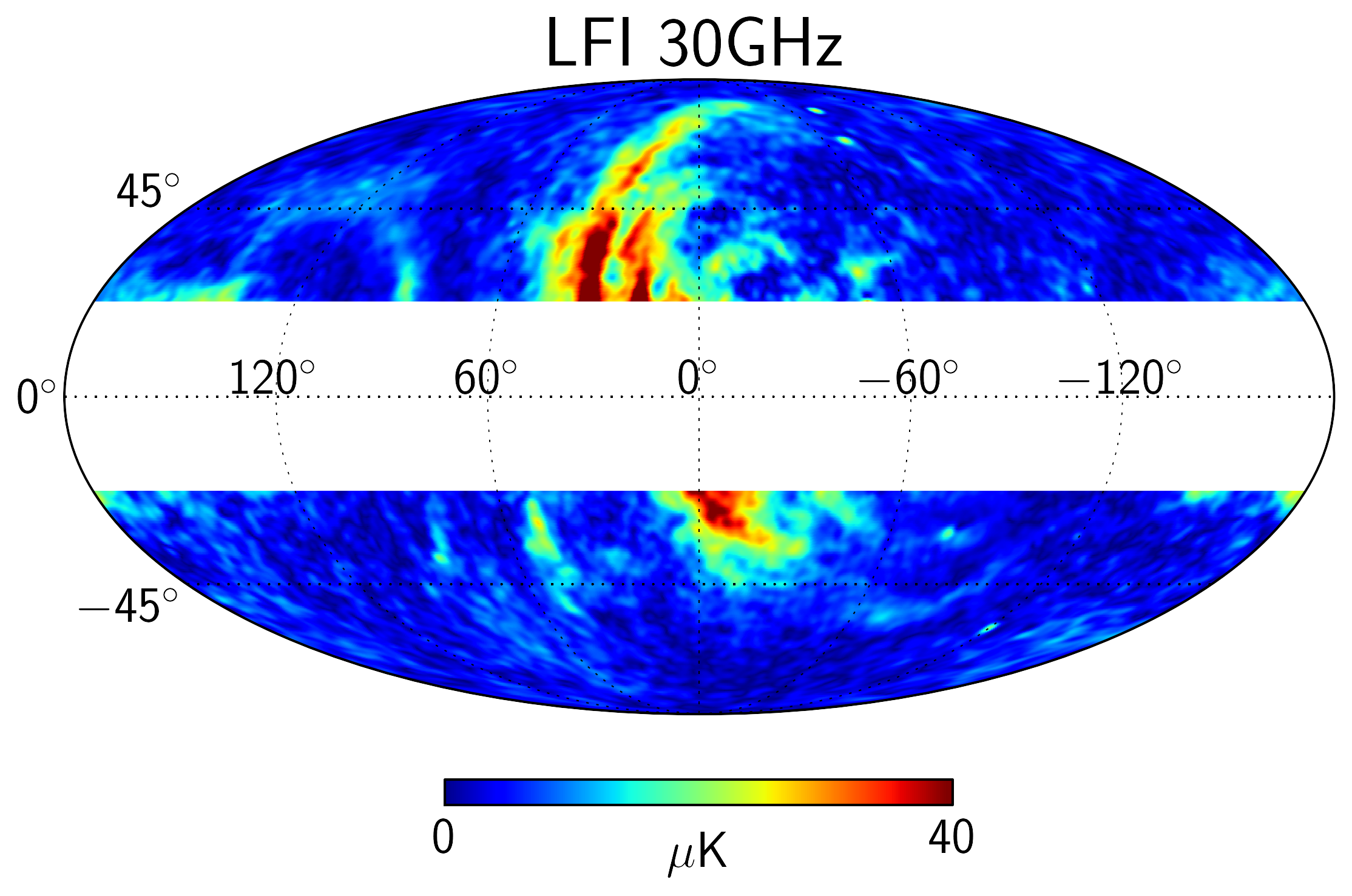}
     \includegraphics[width=8.5 cm]{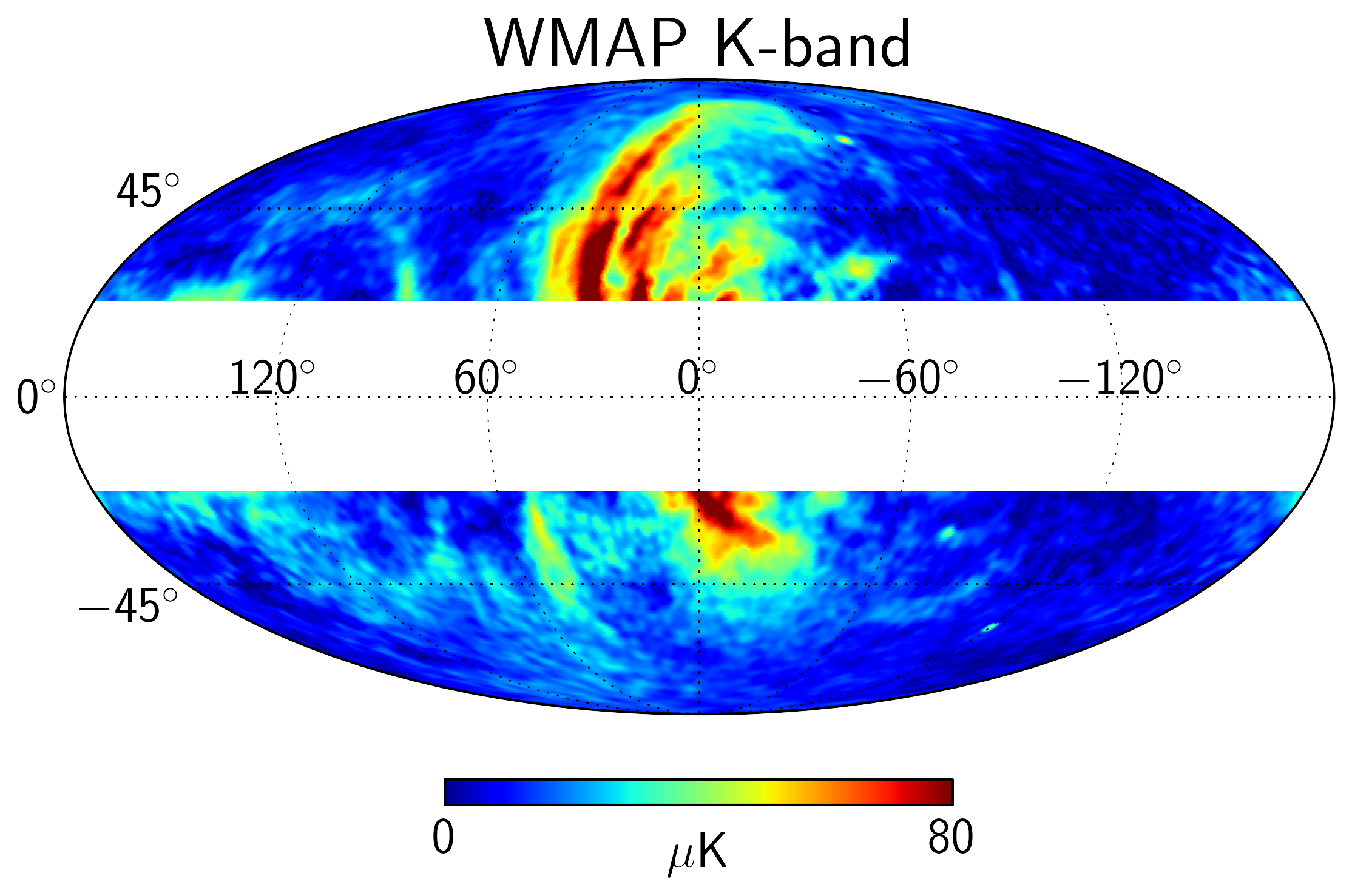}
      \caption{Amplitude of polarized emission ($P=\sqrt{Q^2+U^2}$) for Planck 353, 30 GHz maps (top and middle panels) and WMAP-$K$ band map (bottom panel), in the area considered in this work ($|b|>20^{\circ}$). Maps are in thermodynamical CMB units and have been smoothed to $2^{\circ}$ FWHM.}
         \label{full_maps}
   \end{figure}

\section{Data and methodology}
\label{Sec_data}
\label{Sec_data}
We use the publicly available \textit{Planck} and WAMP-9 years data to get information on the the Galactic thermal dust radiation and synchrotron emission. \par
The \textit{Planck} satellite performed full sky observations in linear polarization in seven different frequency bands (centered at the nominal frequencies of 30, 44, 70, 100, 143, 217 and 353 GHz). 
A description of the full set of maps is reported in \citet{planck2015-II}, \citet{planck2015-VI}, \citet{planck2015-VII} and \citet{planck2015-VIII}. 
WMAP observed the sky in five frequency bands, namely $K$, $Ka$, $Q$, $V$ and $W$, covering a frequency interval from $\sim20$ to $\sim106$ GHz \citep{WMAP2013}. \par 
In particular, we consider the \textit{Planck}-HFI 353 GHz maps
as tracers of the Galactic thermal dust emission. Specifically, we use the time-split ``Half-Mission'' (HM) maps at 353 GHz for estimation of 
power spectra. This pair of maps has highly independent noise and the cross-correlation between them leads to power spectra with negligible noise bias. For synchrotron, we consider the 
\textit{Planck}-LFI 30 GHz and WMAP-$K$ band maps, which are cross-correlated to estimate power spectra. We apply to the maps a point source mask which filters out the polarized sources included in the Second \textit{Planck} 
Catalogue of Compact Sources (PCCS2) at frequency of 30 and 353 GHz \citep{planck2015-XXVI}. \par 
Table \ref{maps_info} reports the main parameters of the used maps, while Figure \ref{full_maps} shows the polarization amplitude maps ($P=\sqrt{Q^2+U^2}$, 
with $Q$ and $U$ being the two Stokes parameters describing the linear polarization state of incoming radiation), at the latitudes of interest here, corresponding to $|b|\ge 20^{\circ}$. 
The maps have been smoothed considering a Gaussian circular beam of 2$^{\circ}$, corresponding to multipole $\ell\lesssim80$, the angular scale at which the CMB GWs
$B$-modes dominate over the GL signal.\par
We characterize the foreground emission by computing polarization power spectra on a set of small size masks, similar to the one used in PIP-XXX: 352 sky regions obtained considering circles with radius $14^{\circ}$ (area of about 600 deg$^{2}$) 
at the pixel centres of the HEALPix $N_{\text{side}}=8$ grid \citep{Gorski05} with Galactic latitude $|b|>35^{\circ}$. Each of the obtained  patch covers a fraction of the sky of about 1.5\%, a typical dimension of observation field for ground-based or balloon-borne experiments. The results of the analysis on these sky patches are presented in Section \ref{Sec:FG_cont}. \par

\subsection{{\tt Xpure} power spectrum estimator}
The computation of angular power spectra on incomplete sky coverage requires the application of specific algorithms in order to correct for the multipole mixing and the polarization state mixing (E-to-B leakage) deriving from the sky cut. 

A variety of procedures have been proposed  and implemented into data analysis 
algorithms which are commonly used, see e.g. the {\tt Xspect} and {\tt Xpol} codes \citep{Tristram05}. 
In this paper, we adopt the {\tt Xpure}\footnote{\url{www.apc.univ-paris7.fr/APC_CS/Recherche/Adamis/MIDAS09/software/pures2hat/pureS2HAT.html}}
methodology, which is a numerical implementation of the pseudo-pure approach described in \citet{Smith06} and \citet{Grain09}. It uses a suitable apodization of the considered sky patch that vanish at the border together with its first derivative. By calculating the spin-wighted windows of the input window function, it estimates the pure-$C_{\ell}$, which ideally are free from the E-to-B leakage. The {\tt Xpure} 
code can handle multiple maps for computing auto and cross power spectra. 
\par
In our analysis we make use of a cosine squared apodization of the sky masks, with an apodization scale equal to $2^{\circ}$ for all the 352 circular sky patches. \par

\begin{table*}
      \caption{Characteristics of the WMAP and \textit{Planck} sky maps used in this analysis \citep{WMAP2013, planck2015-I}}
         \label{maps_info}
     $$ 
           \begin{threeparttable}

         \begin{tabular}{lccc} 
            \hline\hline
            \noalign{\smallskip}\noalign{\smallskip}
           &  \textit{Planck}-353 GHz & \textit{Planck}-30 GHz & WMAP-$K$ band \\
            \noalign{\smallskip}
            \hline
            \noalign{\smallskip}
           Central frequency [GHz] & 353 & 28.4 & 22.4\\
           Beam FWHM [arcmin] & 4.8 & 33.2 & 52.8\\
           Mean Q/U total RMS ($2^{\circ}$, $|b|>20^{\circ}$)\tnote{a}\;\;[$\mu$K]&  49.8 & 6.8 & 16.5\\
           Mean Q/U noise RMS ($2^{\circ}$, $|b|>20^{\circ}$)\tnote{a}\;\;[$\mu$K]&3.3 & 2.0 & 2.2\\
            \noalign{\smallskip}
            \hline
         \end{tabular}
         \begin{tablenotes}
            \item[a] The noise level refers to an angular resolution of $2^{\circ}$, assuming a flat $C_{\ell}$ spectrum.
        \end{tablenotes}
     \end{threeparttable}
     $$ 
   \end{table*}

\begin{figure*}
   \centering
   \includegraphics[width=9 cm]{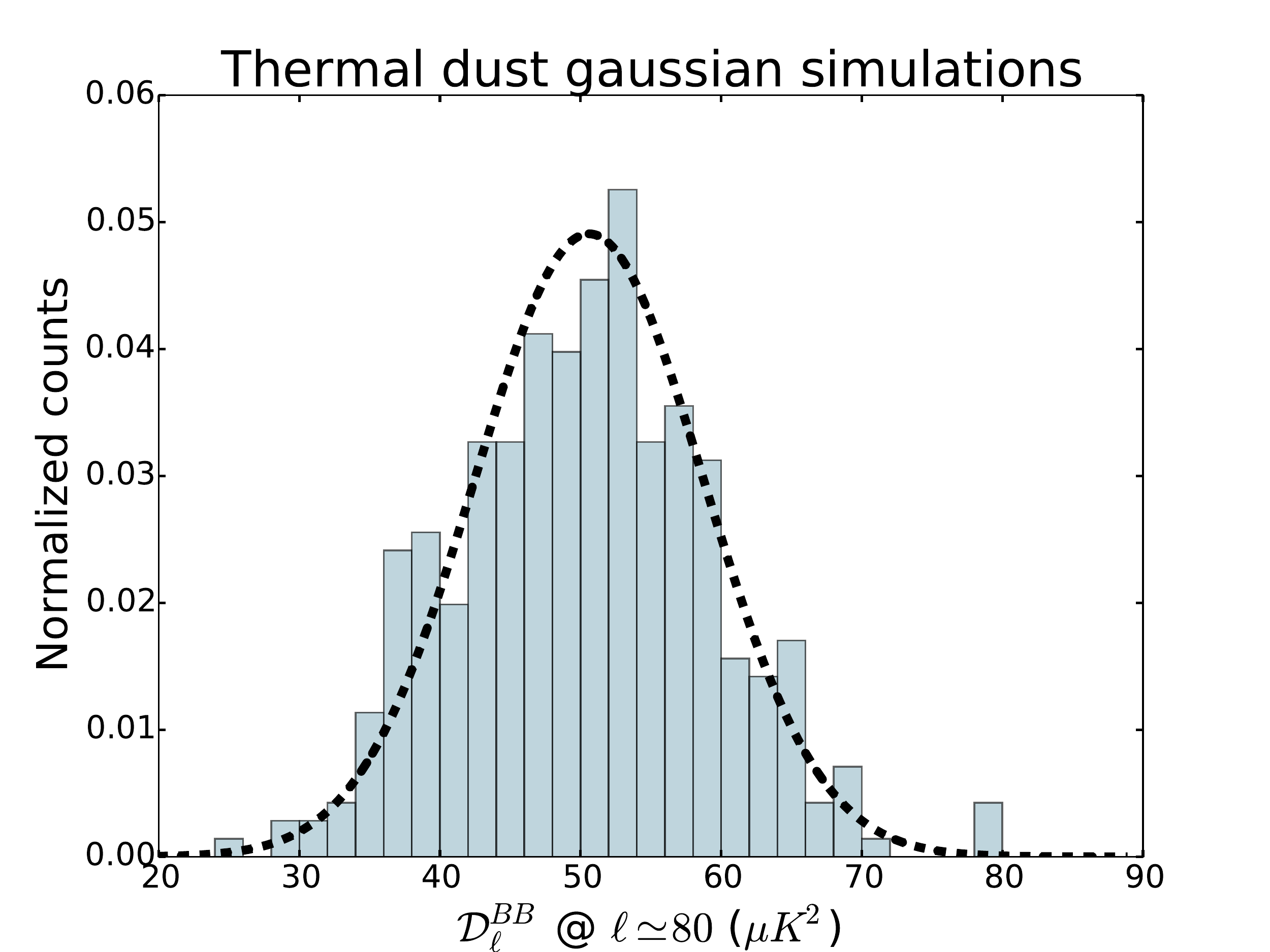}
    \includegraphics[width=9 cm]{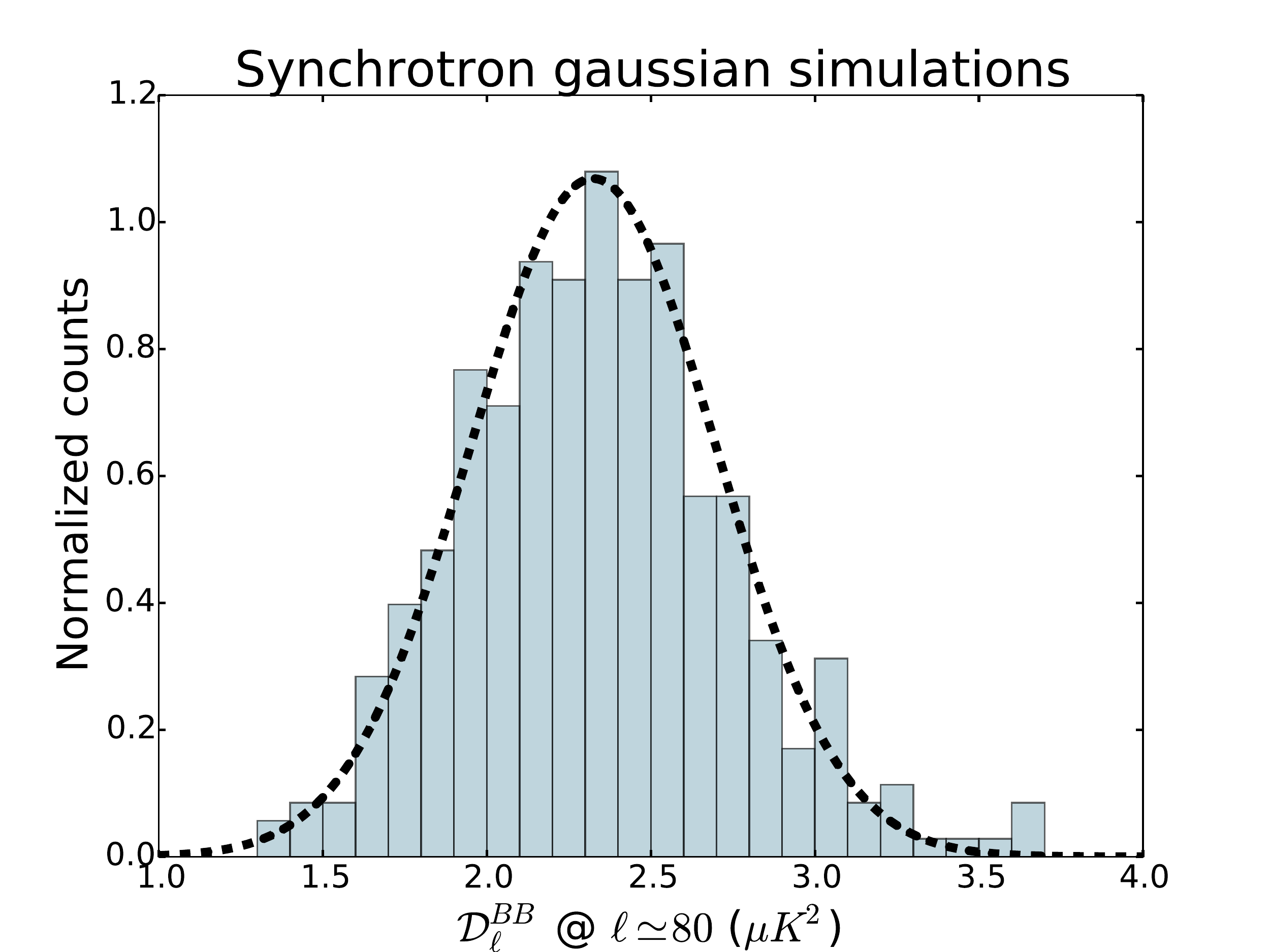}
         \caption{ Validation of the {\tt Xpure} power spectrum estimator for thermal dust (left panel) and synchrotron (right panel) gaussian simulations. Grey histograms are obtained from the $B$-modes power spectra estimation of the input simulated maps in the multipole bin centered at $\ell\simeq80$, in the 352 considered sky regions. Black curves represent the expected gaussian distribution centered on the mean value of the input spectra, from which simulated maps are computed, and with variance defined as in equation (\ref{sample_var}).
              }
         \label{Xpure_val}
   \end{figure*}

\section{Foreground contamination to CMB $B$-modes}

 \begin{figure*}[!t]
   \centering
        \includegraphics[width=8.5 cm]{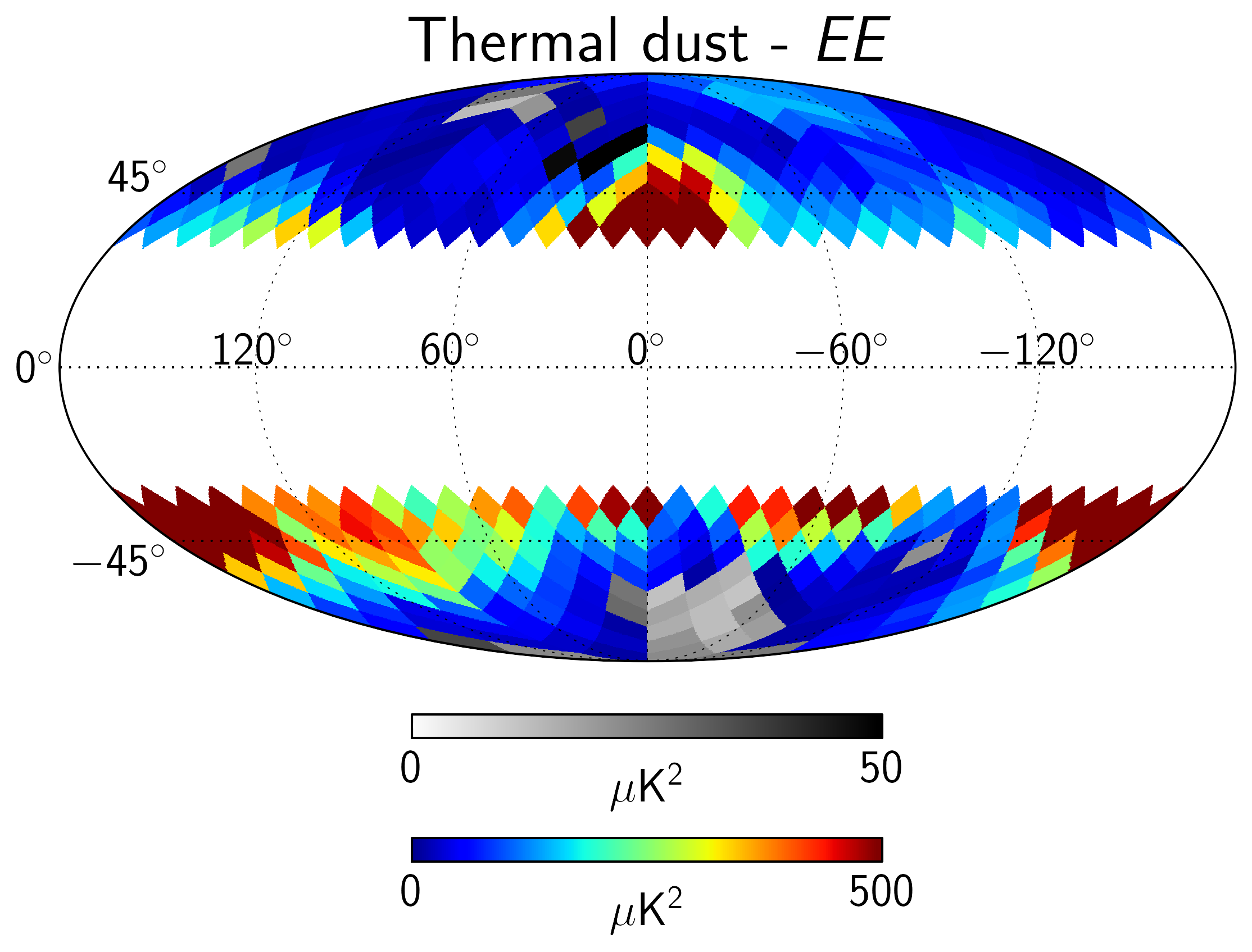}
    \includegraphics[width=8.5 cm]{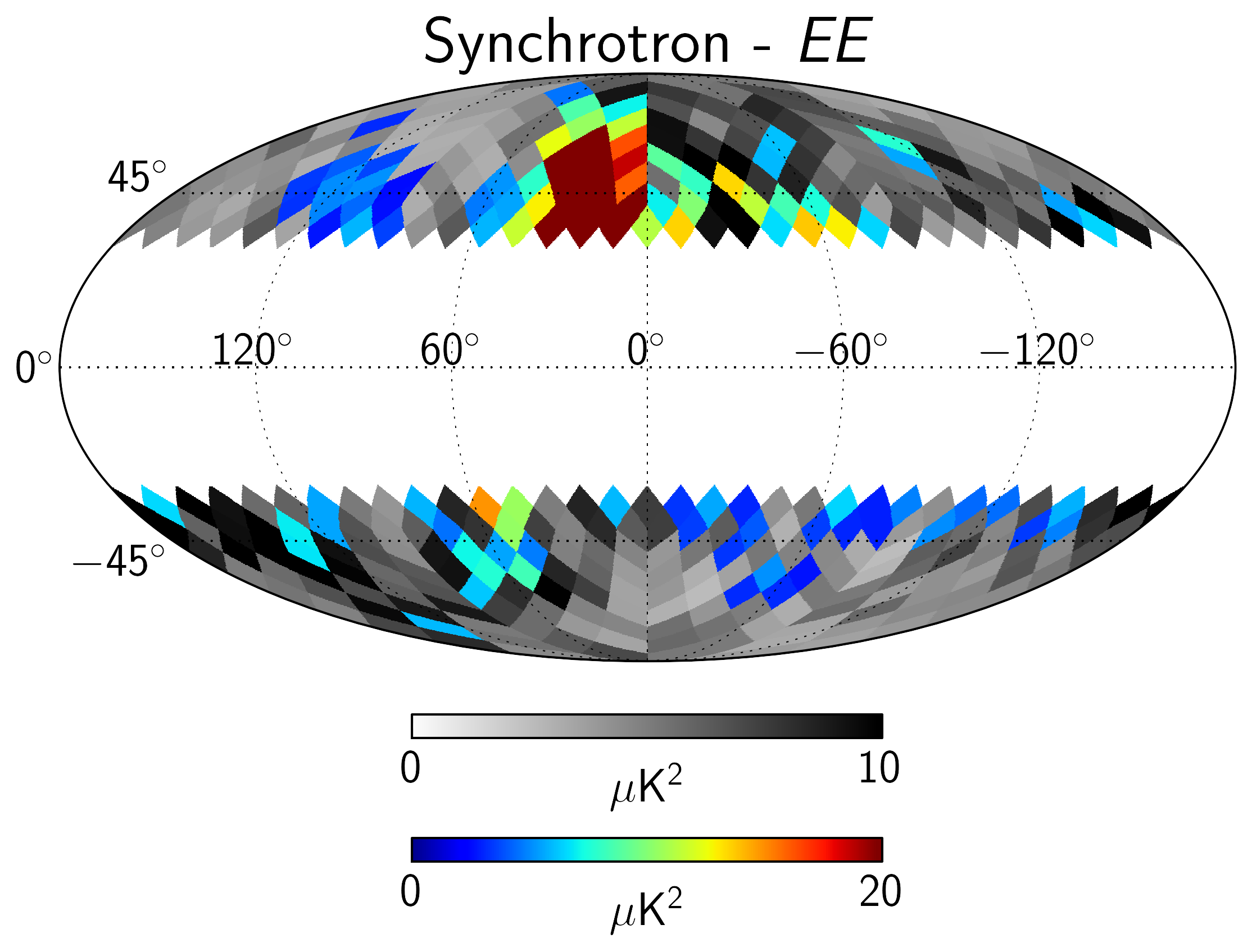}
   \includegraphics[width=8.5 cm]{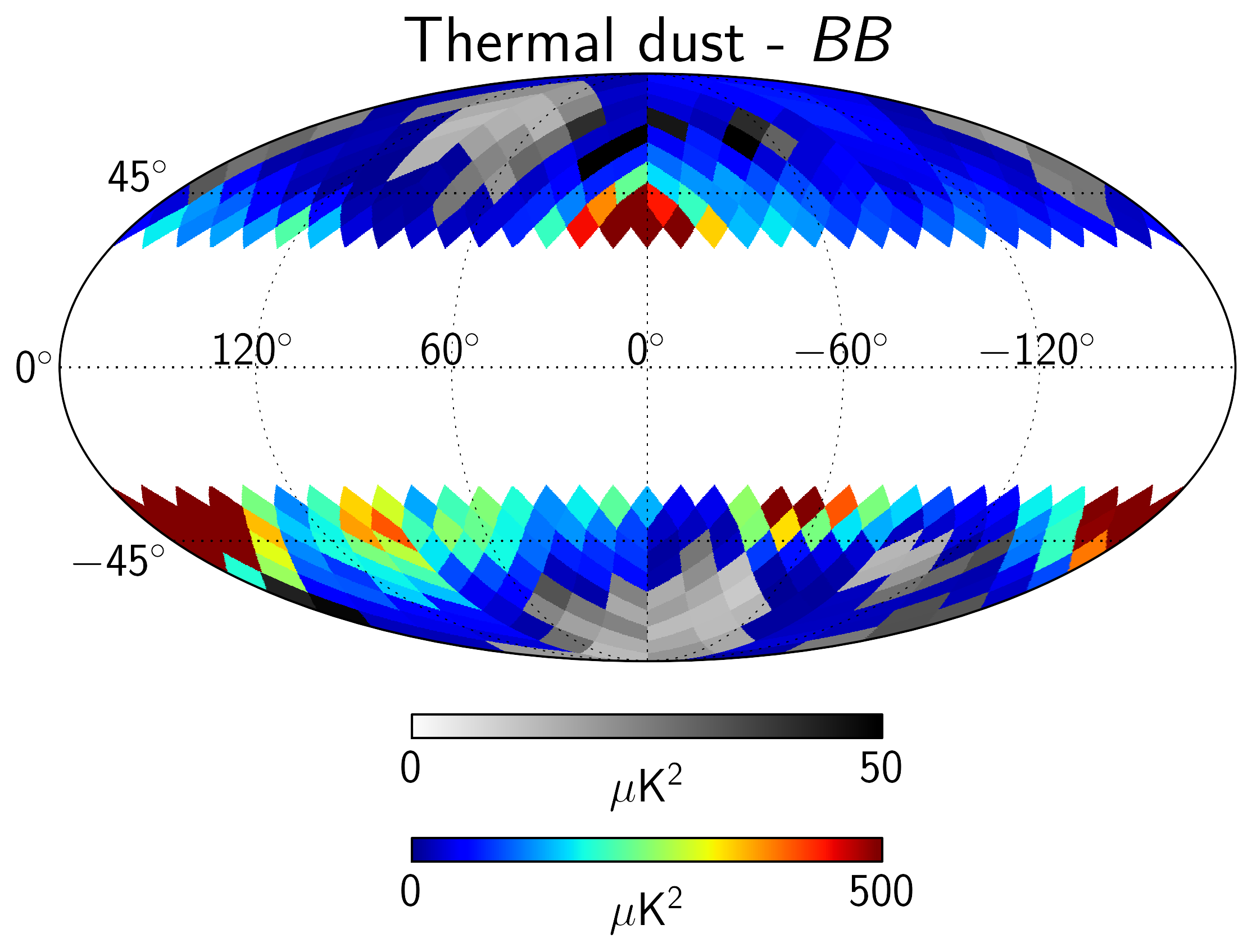}
    \includegraphics[width=8.5 cm]{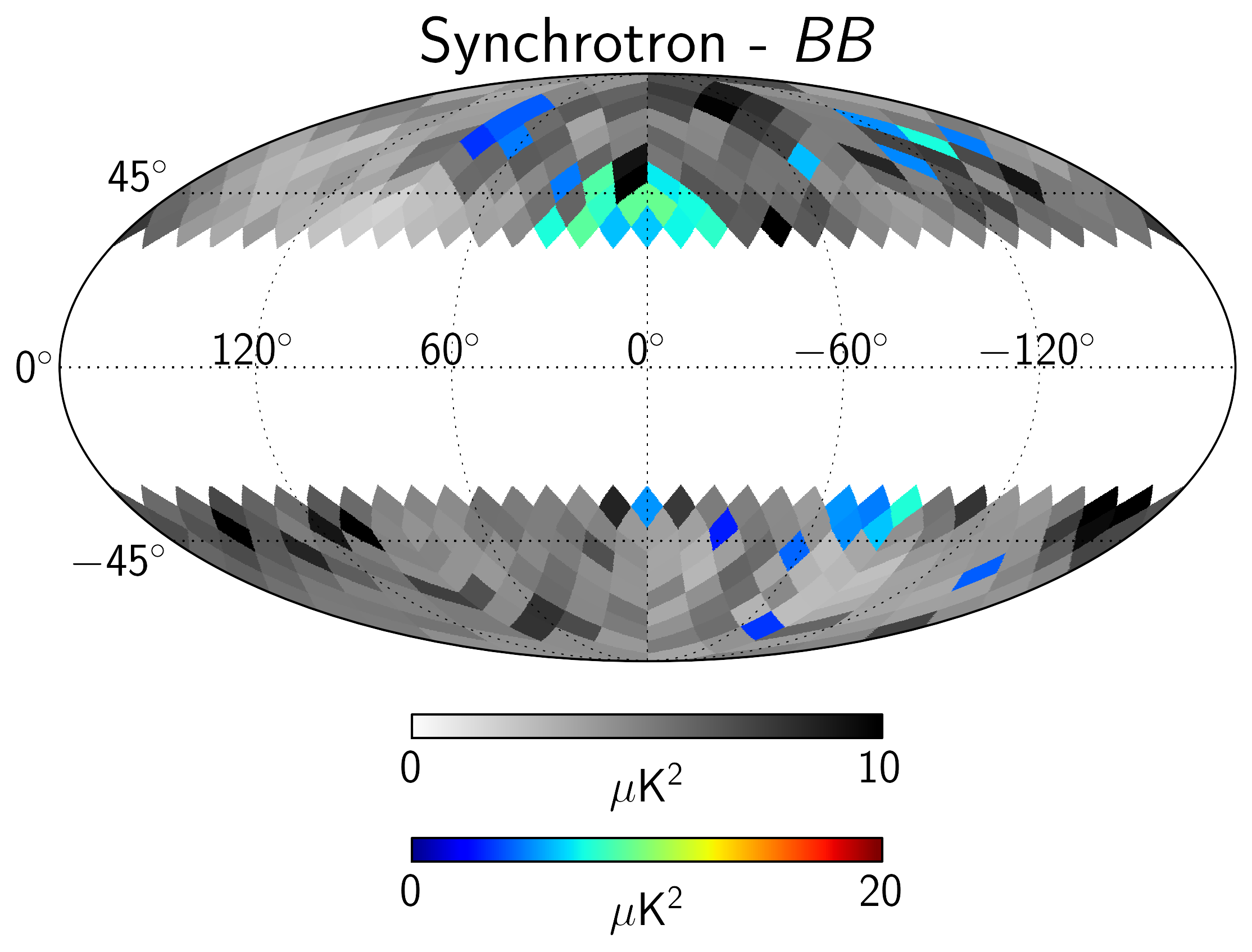}
         \caption{$E$ (top) and $B$-modes (bottom) power spectra measurements in the $\ell_{80}$ multipole bin for dust in the \textit{Planck} 353 GHz channel (left) and synchrotron (right) via cross-correlation of the LFI 30 GHz and WMAP-$K$ band. Colored pixels report 
         detection above $3\sigma$. Grey pixels show upper limits where no detection was possible above the threshold, as defined in (\ref{eq_uplim}).}
         \label{l80_maps}
   \end{figure*}

\label{Sec:FG_cont}

In this Section we describe the results on the estimation of the contamination arising from foreground emission on CMB $B$-modes observations. In order to asses this level of contamination we compute power spectra
of the foreground emissions on the set of 352 circular sky masks described previously. Power spectra are computed with {\tt Xpure} on these regions, specifically in the $\ell$-bins defined by the intervals between multipoles: 
40, 60, 100, 140, 180, 220. Nevertheless, since we are interested in the contamination to CMB measurements of primordial tensor modes at the degree angular scale, we only take into account the amplitude of foreground emission recovered in the multipole bin centered at $\ell\simeq80$, hereafter referred to as $\ell_{80}$, where the so-called CMB $B$-modes ``recombination bump'' peaks and where a number of current and forthcoming sub-orbital experiment are focusing to constrain $r$. \par
We recall that we estimate thermal dust spectra by cross-correlating the HFI 353 GHz HM maps, while for synchrotron we cross-correlate the LFI 30 GHz and WMAP-$K$ maps, therefore evaluating the synchrotron polarization amplitude at the effective frequency  $\nu_{\text{eff}}=\sqrt{\nu_{LFI-30}\times\nu_{WMAP-K}}\simeq25.4$ GHz.

\subsection{{\tt Xpure} validation}
\label{Sec:Xpure_val}
We validate the performance of the {\tt Xpure} algorithm, on our analysis setup, by computing power spectra of simulated polarization maps. We check whether the spectra amplitude estimation in the $\ell_{80}$ bandpower, calculated in such small sky patches, is unbiased and with minimum variance. Simulated maps are generated considering Gaussian realizations of input theoretical power spectra in the form $\mathcal{D}_{\ell}^{EE/BB}=A^{EE/BB}\ell^{\alpha+2}$ (where $\mathcal{D}_{\ell}^{EE/BB}\equiv\ell(\ell+1)C^{EE/BB}_{\ell}/2\pi$). This kind of dependence, for  polarization foreground power spectra, has found to be valid for both  dust emission, with  $\alpha\simeq-2.42$ and $(A^{BB}/A^{EE})\simeq0.52$ \citep{planck2014-XXX}, and synchrotron radiation, with $\alpha\simeq-2.37$ and $(A^{BB}/A^{EE})\simeq0.36$ \citep{planck2015-X}.  We, therefore, explore these two cases in our simulations. Even though the hypothesis of Gaussianity does not apply to the Galactic emissions, these simulations give us the possibility to construct maps with known expected mean value and variance of the power spectra in a given multipole bin, to be compared with the values 
retrieved by {\tt Xpure}.\par 
Therefore we calculate power spectra for the two different cases: using maps simulating a Gaussian thermal dust signal having an input power spectrum taking the value $\mathcal{D}^{BB}_{\ell_{80}}=50$ $\mu$K$^{2}$, and maps simulating a Gaussian synchrotron signal with  $\mathcal{D}^{BB}_{\ell_{80}}=2.3$ $\mu$K$^{2}$. These numbers represent the median values of the actual $B$-modes amplitude of dust and synchrotron spectra computed from  \textit{Planck} and WMAP data\footnote{Median values are computed considering only those regions where $\mathcal{D}^{BB}_{\ell_{80}}>0$.} (see Section \ref{Sec_spectra_l80}). We simulate 352 maps for each case, and then compute power spectra on our set of 352 regions, using therefore a different simulated map for each region. Figure \ref{Xpure_val} shows the results of the validation, with the histogram of the {\tt Xpure} estimation of the $B$-modes amplitude in the 352 sky regions in the $\ell_{80}$ bin, 
compared with the Gaussian distribution having mean equal to the value of the input spectrum in the same $\ell$-bin and as variance the approximation of the signal sample variance defined as:
\begin{equation}
\centering
\text{var}(\mathcal{D}^{BB}_{\ell_{\text{bin}}})=\frac{2}{(2\ell_{\text{bin}}+1)f^{\text{eff}}_{\text{sky}}\,\Delta\ell_{\text{bin}}}(\mathcal{D}^{BB}_{\ell_{\text{bin}}})^{2},
\label{sample_var}
\end{equation}
where $f^{\text{eff}}_{\text{sky}}$ is the mean effective sky fraction retained by the masks, which is approximately\footnote{The retained effective sky fraction is not exactly the same for each region, since compact sources are masked, and varies between 0.0115 and 0.0126} 0.0125.\par
In both cases the histogram recovered from data agrees with the expected Gaussian distribution, with $p$-values obtained from a statistical null-hypothesis test equal to 0.48 and 0.52 for thermal dust and synchrotron case 
respectively\footnote{$p$-value means the probability to get the observed distribution or a ``more unlikely'' one if data are indeed coming from the previously defined Gaussian probability function} (the analysis on $E$-modes spectra gives similar results). From this validation 
we can therefore conclude that {\tt Xpure} estimates correctly the power spectra amplitude in the $\ell_{80}$ bandpower in the considered sky patches. 

\subsection{Foreground amplitude in the $\ell_{80}$ bandpower}
\label{Sec_spectra_l80}

\begin{table*}
      \caption{Definition of classes for sky regions.}
         \label{r_min_class}
     $$ 
         \begin{tabular}{rlcc} 
            \hline\hline
            \noalign{\smallskip}\noalign{\smallskip}
           \multicolumn{2}{l}{Class} &  Number of regions & Color scale on Figure \ref{freq_r_min} \\
            \noalign{\smallskip}
            \hline
            \noalign{\smallskip}
           1.& Detection of thermal dust and synchrotron & 28 & Green\\
           2.& Detection of thermal dust, upper limit on synchrotron & 250 & Grey\\
           3.& Detection of synchrotron, upper limit on thermal dust & 4 & Blue\\
           4.&Upper limit on thermal dust and synchrotron & 70 & Red\\
            \noalign{\smallskip}
            \hline
         \end{tabular}
     $$ 
      \end{table*}

Maps in Figure \ref{l80_maps} are constructed from the {\tt Xpure} evaluation of power spectra at $\ell\simeq80$ of \textit{Planck} and WMAP maps. In particular, for both the HFI$_{353}$ HM$_1\times$HM$_2$ and LFI$_{30}\times$WMAP$_{K}$ cases, each pixel shows the value of the $E$ or $B$-modes power spectra in the $\ell_{80}$ bandpower, computed in the corresponding circular region. The maps show the power spectrum amplitude for those pixel where there is a positive detection of signal at 3$\sigma$ 
(colored scale on maps). Errors are obtained from one hundred simulations of white noise maps generated from the $Q$ and $U$ pixel-pixel covariance matrices of the input sky maps. In all  cases where the power spectrum 
at $\ell\simeq80$ does not exceed $3\sigma$, the maps report  upper limits (grey scale on maps). Upper limits are defined in the following conservative way: 
\begin{equation}
\begin{aligned} 
-\;\; {\rm if}\ &\mathcal{D}^{BB/EE}_{\ell_{80}}>0:\ {\rm UL}(\mathcal{D}^{BB/EE}_{\ell_{80}})=\mathcal{D}^{BB/EE}_{\ell_{80}}+3\sigma(\mathcal{D}^{BB/EE}_{\ell_{80}})\,\\ 
-\;\;{\rm if}\ &\mathcal{D}^{BB/EE}_{\ell_{80}}<0:\ {\rm UL}(\mathcal{D}^{BB/EE}_{\ell_{80}})=3\sigma(\mathcal{D}^{BB/EE}_{\ell_{80}})\ .
\label{eq_uplim}
\end{aligned} 
\end{equation}
\par
For both thermal dust and synchrotron the number of positive detections is larger for $E$ than $B$-modes, as expected given the $E$/$B$ asymmetry of the two kinds of emission \citep{planck2014-XXX, planck2015-X}. For thermal dust positive detections are obtained in 325 and 278 regions for $E$ and $B$-modes respectively, with values ranging from $\sim10$ to more than 2500 $\mu$K$^2$ (errors between $\sim2$ and $\sim10$ $\mu$K$^2$).  Upper limits range between $\sim10$ and $\sim60$ $\mu$K$^2$. For $B$-modes the recovered amplitudes are consistent with the similar ones reported in PIP-XXX, even though the circular regions we consider here are are slightly larger, with discrepancies above $3\sigma$ only in 10 sky patches.\par 
The situation for synchrotron is more critical, because of the lower signal-to-noise ratio of the data. Positive detections are found only in 101 regions for $E$-modes and 32 regions for $B$-modes, with amplitudes between $\sim3$ and $\sim40$  $\mu$K$^2$ (errors between $\sim0.6$ and $\sim3$  $\mu$K$^2$). Upper limits range between $\sim2$ and $\sim10$  $\mu$K$^2$. The number of regions with positive detection for synchrotron $B$-modes increases to 66 or 114 if we consider threshold at $2\sigma$ or $1\sigma$ respectively.\par 
The morphology of the retrieved maps follows the morphology of the foreground emissions, with larger amplitude in the $\ell_{80}$ bandpower in those regions where the total polarized signal is stronger (compare with Figure \ref{full_maps}). 

\subsection{Looking for the minimum of foreground emission}

  \begin{figure*}
   \centering
   \includegraphics[width=16 cm]{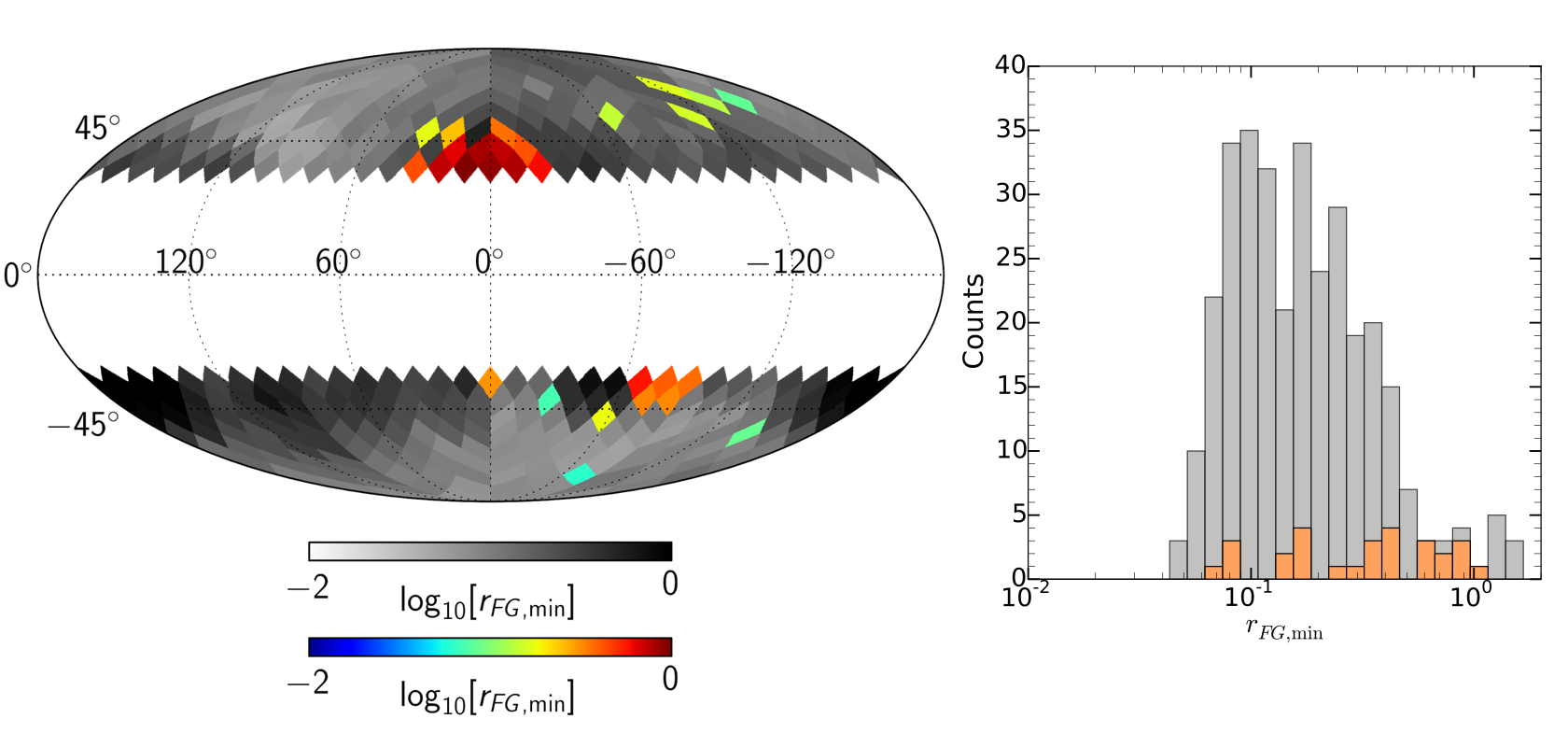}\\
    \includegraphics[width=16 cm]{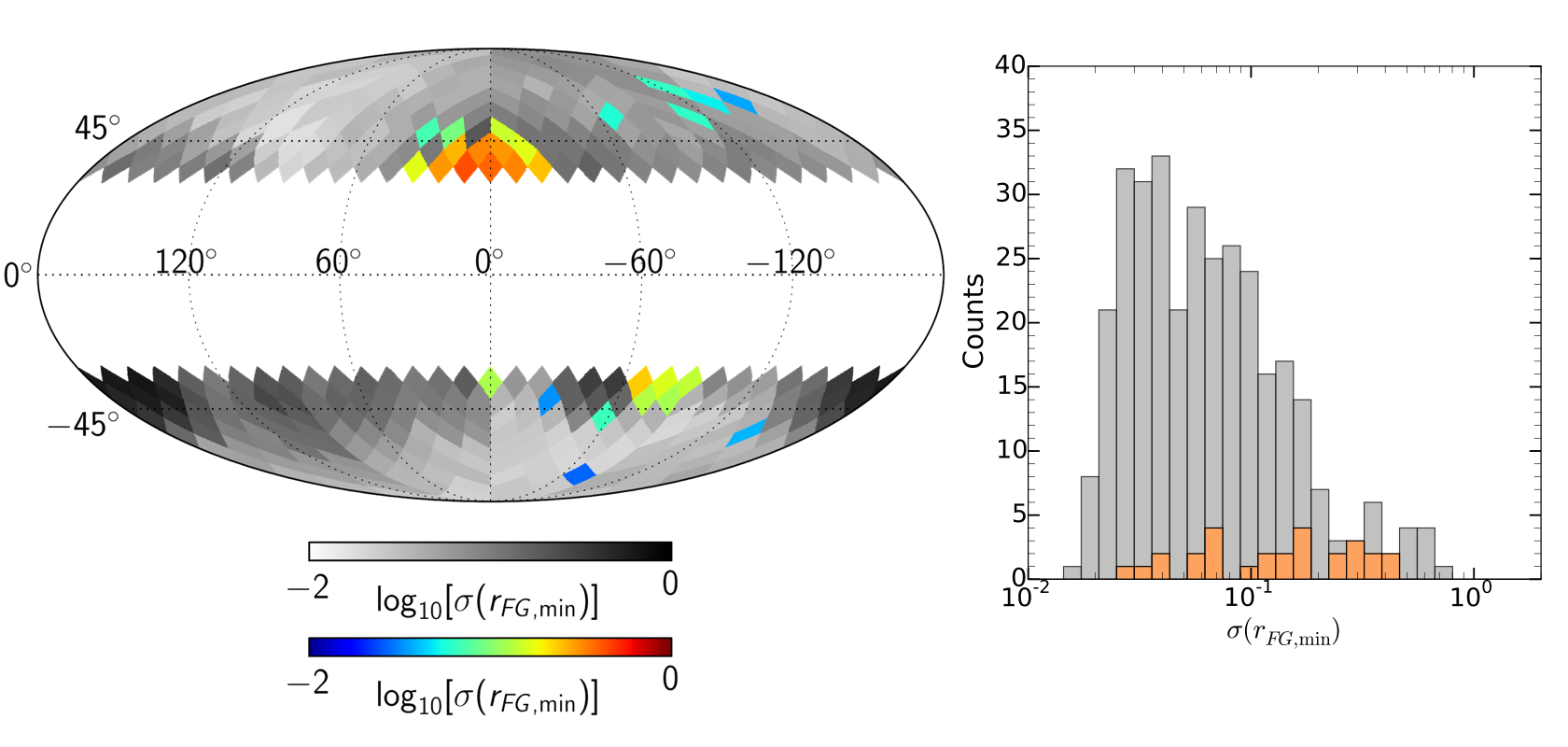}
     
         \caption{Minimum estimated foreground emissions ($r_{FG, \text{min}}$) in units of the cosmological tensor-to-scalar ratio (top) and uncertainties on it (bottom). On maps, colored pixels refer to detection, grey pixels to upper limits. Histograms are obtained from the retrieved values reported on maps, orange bars for detection, grey bars for upper limits.}
         \label{r_min}
   \end{figure*}
 
  \begin{figure*}[!t]
   \centering
    \includegraphics[width=16 cm]{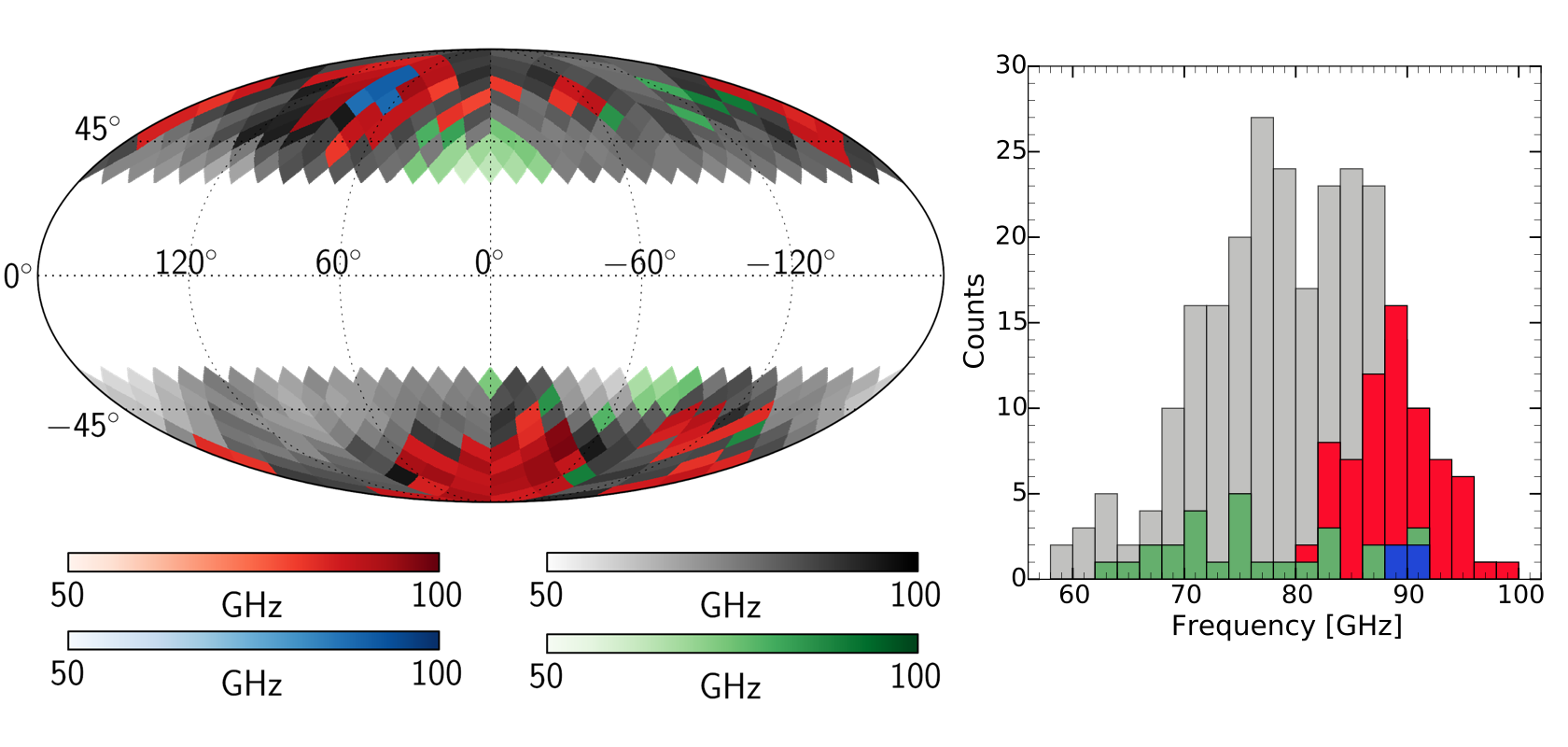}
             \caption{Map and histogram reporting the frequency at which the minimum of foreground emission is found in each sky region. Color codes as described in Table \ref{r_min_class}.}.
         
         \label{freq_r_min}
   \end{figure*}

 \begin{table*}
      \caption{Summary of the characteristics of regions of interests: patches where both synchrotron and thermal dust are detected in the $\ell_{80}$ bandpower and with recovered value for $r_{FG, \text{min}}<0.12$ (first four rows) and region where the upper limit on $r_{FG, \text{min}}$ is below $\sim0.06$ (remaining sixteen entries).
      From left to right: HEALPix pixel number, direction of the centre, $r_{FG, \text{min}}$, 
      frequency of minimum emission, frequencies corresponding to synchrotron and dust being negligible within the quoted threshold, region class.}
         \label{r_min_info}
     $$ 
       \begin{threeparttable}
         \begin{tabular}{ccccccc} 
            \hline\hline
            \noalign{\smallskip}\noalign{\smallskip}
           Pixel number\tnote{a} & Coordinates\tnote{b} & $r_{FG, \text{min}}$ &   $r_{FG, \text{min}}$& $r_s<0.005$ & $r_d<0.005$ & Class\tnote{c} \\           
            \noalign{\smallskip}
                                               &  $(b, l)$        &                                  & Freq. [GHz]        & Freq. [GHz]    & Freq. [GHz] &        \\
             \noalign{\smallskip}                                   
            \hline
            \noalign{\smallskip}
             51\dotfill & $(60.43^{\circ}, -153.00^{\circ})$ & $0.09\pm0.04$ & $91\pm7$ & $135\pm9$& $46\pm11$ &1\\
             653\dotfill & $(-41.81^{\circ}, -28.12^{\circ})$ & $0.07\pm0.03$ & $86\pm7$& $121\pm9$ & $45\pm10$ &1\\
             698\dotfill & $(-54.34^{\circ}, -142.50^{\circ})$ & $0.09\pm0.04$ & $88\pm7$& $130\pm10$ & $44\pm11$ &1\\
             753\dotfill & $(-72.39^{\circ}, -75.00^{\circ})$ & $0.06\pm0.03$ & $90\pm7$& $126\pm8$ & $50\pm11$&1\\ 
            \noalign{\smallskip}
            \hline
            \noalign{\smallskip}
             29\dotfill & $(66.44^{\circ}, 123.75^{\circ})$ &  $0.058\pm0.021$ &  $94\pm5$ & $129\pm4$& $54\pm12$ & 2\\
             46\dotfill & $(60.43^{\circ}, 117.00^{\circ})$ &  $0.061\pm0.022$ &  $91\pm4$ & $125\pm4$ &$51\pm11$ & 2\\
             48\dotfill & $(60.43^{\circ}, 153.00^{\circ})$ &  $0.064\pm0.024$ &  $90\pm5$ &$124\pm3$ &$50\pm11$& 2\\
             67\dotfill & $(54.34^{\circ}, 112.50^{\circ})$ &  $0.048\pm0.017$ &  $93\pm4$ & $122\pm4$ &$57\pm12$& 2\\
             90\dotfill & $(48.14^{\circ}, 83.57^{\circ})$ &  $0.064\pm0.023$ &  $87\pm5$ &$120\pm3$ & $48\pm11$& 2\\
             91\dotfill & $(48.14^{\circ}, 96.43^{\circ})$ &  $0.051\pm0.018$ &  $94\pm5$ &$124\pm4$&$56\pm12$& 2\\
             92\dotfill & $(48.14^{\circ}, 109.29^{\circ})$ &  $0.050\pm0.017$ &  $94\pm5$ &$125\pm4$&$57\pm11$& 2\\
             93\dotfill & $(48.14^{\circ}, 122.14^{\circ})$ &  $0.059\pm0.021$ &  $94\pm5$ & $129\pm4$ &$54\pm11$& 2\\
             119\dotfill & $(41.81^{\circ}, 84.37^{\circ})$ & $0.055\pm0.020$  &  $87\pm4$ & $116\pm3$ &$50\pm11$& 2\\
             120\dotfill & $(41.81^{\circ}, 95.62^{\circ})$ & $0.052\pm0.019$  &  $83\pm5$ &$109\pm3$  &$49\pm11$& 2\\
             152\dotfill & $(35.69^{\circ}, 90.00^{\circ})$ & $0.060\pm0.022$  &  $84\pm5$ & $113\pm3$  &$47\pm10$& 2\\
             701\dotfill & $(-54.34^{\circ}, -97.50^{\circ})$ &  $0.060\pm0.020$ &  $86\pm4$ & $117\pm3$  &$49\pm11$& 4\\
             723\dotfill & $(-60.43^{\circ}, -81.00^{\circ})$ & $0.055\pm0.020$  &  $88\pm5$ &$118\pm3$  &$51\pm11$& 2\\
             724\dotfill & $(-60.45^{\circ}, -63.00^{\circ})$ &  $0.064\pm0.024$ &  $95\pm5$ &$133\pm4$ &$53\pm12$& 2\\
             740\dotfill & $(-66.44^{\circ}, -78.75^{\circ})$ &  $0.061\pm0.021$ &  $88\pm5$ &$118\pm3$&$50\pm11$& 2\\
             741\dotfill & $(-66.44^{\circ}, -56.25^{\circ})$ &  $0.064\pm0.028$ &  $89\pm5$ &$121\pm3$&$50\pm10$& 4\\
                    \noalign{\smallskip}
            \hline
         \end{tabular}

            \begin{tablenotes}
            \item[a] Pixel number of the HEALPix $N_{\text{side}}=8$ grid (enumeration starts from zero, ring ordering) associated to the circular region on which spectra are evaulated.
            \item[b] Galactic coordinates of the region center.
             \item[c] Region class as defined in Table \ref{r_min_class}.
        \end{tablenotes}
     \end{threeparttable}
     $$ 
      \end{table*}

We now use the results described above to evaluate, for each sky region, the expected contamination arising from foregrounds to CMB $B$-modes measurements. Our goal is to estimate, for each region, the frequency at which the total polarized foreground emission reaches its minimum and its corresponding amplitude. To achieve this goal, we extrapolate in frequency the $BB$ maps of Figure \ref{l80_maps}. \par
The frequency scaling is computed using the available information on the Spectral Energy Distribution (SED) of thermal dust and synchrotron radiations 
in polarization.
For thermal dust, we adopt a modified blackbody with $\beta_d=1.59\pm0.17$ at $T_d=19.6$ \citep{planck2014-XXII, planck2014-XXX}. We rescale the synchrotron radiation considering a power law with spectral index $\beta_s=-3.12\pm0.04$, as reported in \citet{Fuskeland14} for polarized emission at intermediate and high Galactic latitudes. In extrapolating the foreground contribution at the various frequencies we take into account the \textit{Planck} 
353 and 30 GHz and WMAP-$K$ band color corrections to properly include the real frequency response of the instruments \citep{planck2013-IX, planck2015-V}.\par 

We recall that, synchrotron and dust $BB$ maps of Figure \ref{l80_maps} report the actual value of the spectra in the $\ell_{80}$ bandpower, in those pixels corresponding to regions where we have a signal detection above $3\sigma$, and the upper limit on it, in all the other pixels. Nevertheless, we treat detections and upper limits in the same way, by rigidly extrapolating in frequency the entire maps.  To distinguish among different cases, we divide the sky regions in four  classes, depending on whether we reach detections or upper limits for synchrotron and dust, and, on the following figures, we use different colors to distinguish them. Classes are 
defined in Table \ref{r_min_class}. \par
Figure \ref{r_min} shows  the minimum foreground amplitude (sum of synchrotron and dust contributions) recovered in each sky patch after frequency extrapolation, together with the associated error.  Amplitudes 
are expressed in terms of $r_{FG, \text{min}}$, computed dividing the total $\mathcal{D}_{\ell}^{BB}$ minimum amplitude of foreground emission in $\mu$K$^2$ by the value of the CMB primordial $B$-modes power spectrum with tensor-to-scalar ratio $r=1$ in the $\ell_{80}$ bandpower, which is equal to $\sim6.67\times10^{-2}$ $\mu$K$^2$. Therefore, for example, a value of $r_{FG, \text{min}} = 0.1$ means a foreground contribution at the level of a CMB GWs signal with $r= 0.1$.\par
For regions belonging to class (1), maps in Figure \ref{r_min} report, with colored pixels, the estimated values of $r_{FG, \text{min}}$ and the associated errors. For each map we also report the histogram of the retrieved values. Uncertainties are calculated as the quadratic sum of the statistical errors, coming from the white noise in the sky maps, and the errors coming from the uncertainties on the frequency scaling parameters. For regions in classes  (2), (3) and (4), we are able to put only upper limits on $r_{FG, \text{min}}$, which are shown with grey pixels in Figure  \ref{r_min}.  We also report the errors on these upper limits which, as before, are the quadratic sum of the statistical errors (included only if we have a detection of at least one of the two kinds of signal) and the frequency scaling ones.\par 
Figure \ref{freq_r_min} shows the map reporting, for each region, the frequency at which we find the minimum of foreground emission, together with the associated histogram. On this map, we again distinguish the four classes of regions using different color scales. For class (1), since we are dealing with detections, the value of each pixel represents the actual frequency where the minimum of foreground emission is reached. For class (2) and (3) the reported value represents the frequency at which we set the upper limit on $r_{FG, \text{min}}$, in the corresponding region. This frequency should be seen as an upper (for class (2)) or a lower limit (for class (3)) on the real frequency at which the foreground emission reaches its actual minimum. For class number (4) each pixel value again represent the frequency corresponding to the upper limit on $r_{FG, \text{min}}$, but, in this case, we are not able to say whether the real frequency of the foreground minimum lies above or below the reported value. \par
The recovered amplitudes of $r_{FG, \text{min}}$, considering both detections and upper limits, vary between $0.05\pm0.02$ and $1.5\pm0.7$ (between $0.06\pm0.03$ and $1.0\pm0.4$ for detections only)  in the frequency range 58-98 GHz (with errors between 4 and 7 GHz). In 123 regions we find $r_{FG, \text{min}}<0.12$, the 
current upper limit on the CMB tensor-to-scalar ratio from $B$-modes observations \citep{B2P}. Table \ref{r_min_info} summarizes the results obtained on several regions of interests. The first four regions correspond to those where
both synchrotron and dust are detected and  $r_{FG, \text{min}}<0.12$. The remaining ones are the regions where we are able to put the most stringent upper limits on  $r_{FG, \text{min}}$, having   $r_{FG, \text{min}}\lesssim0.06$.\par
Within the uncertainties of our analysis, we can conclude that there is no region in the sky where the foreground emission demonstrates to contaminate the CMB $B$-modes at level lower than a signal with tensor-to-scalar ratio $r\sim0.05$. 

           \begin{figure*}
   \centering
   \includegraphics[width=16 cm]{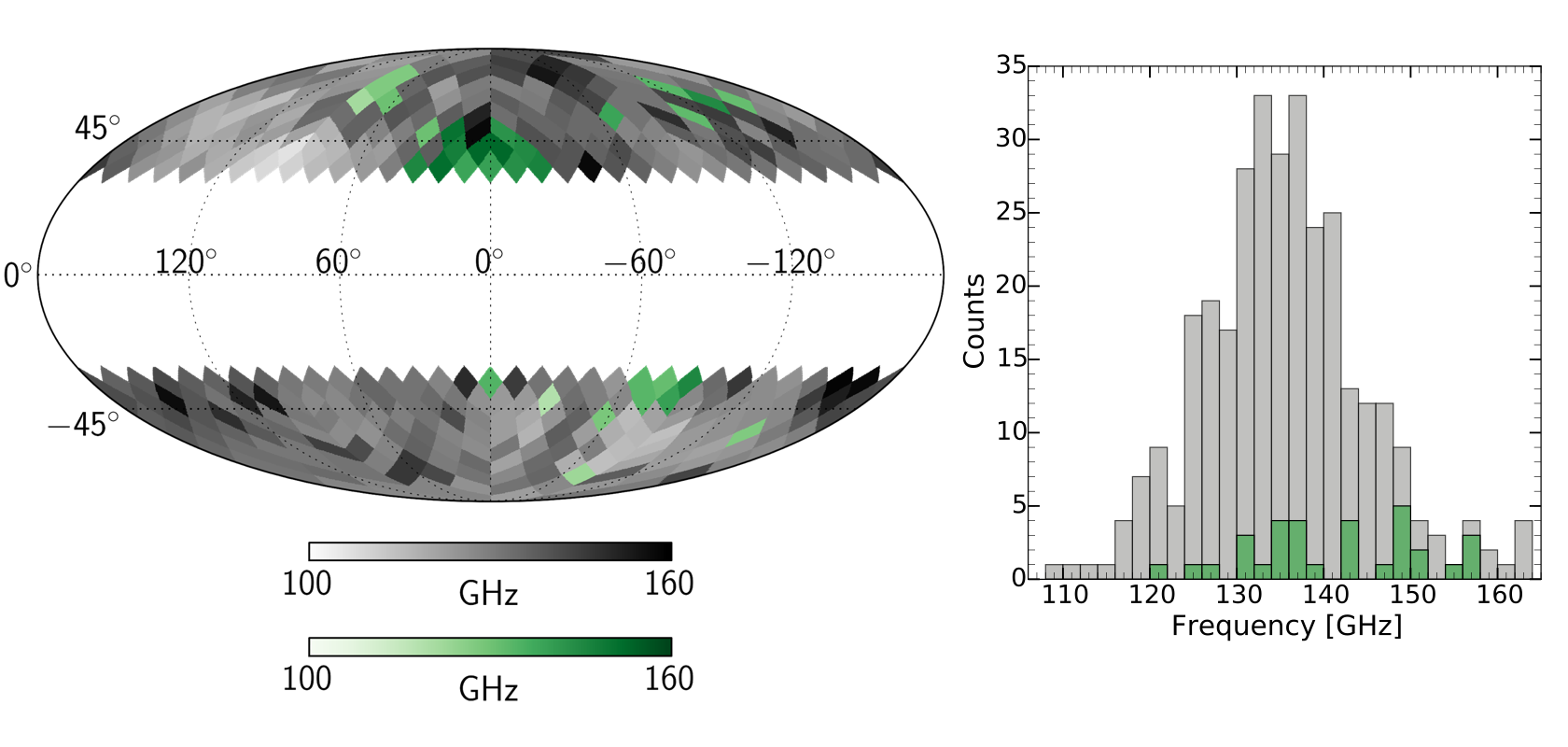}\\
     \includegraphics[width=16 cm]{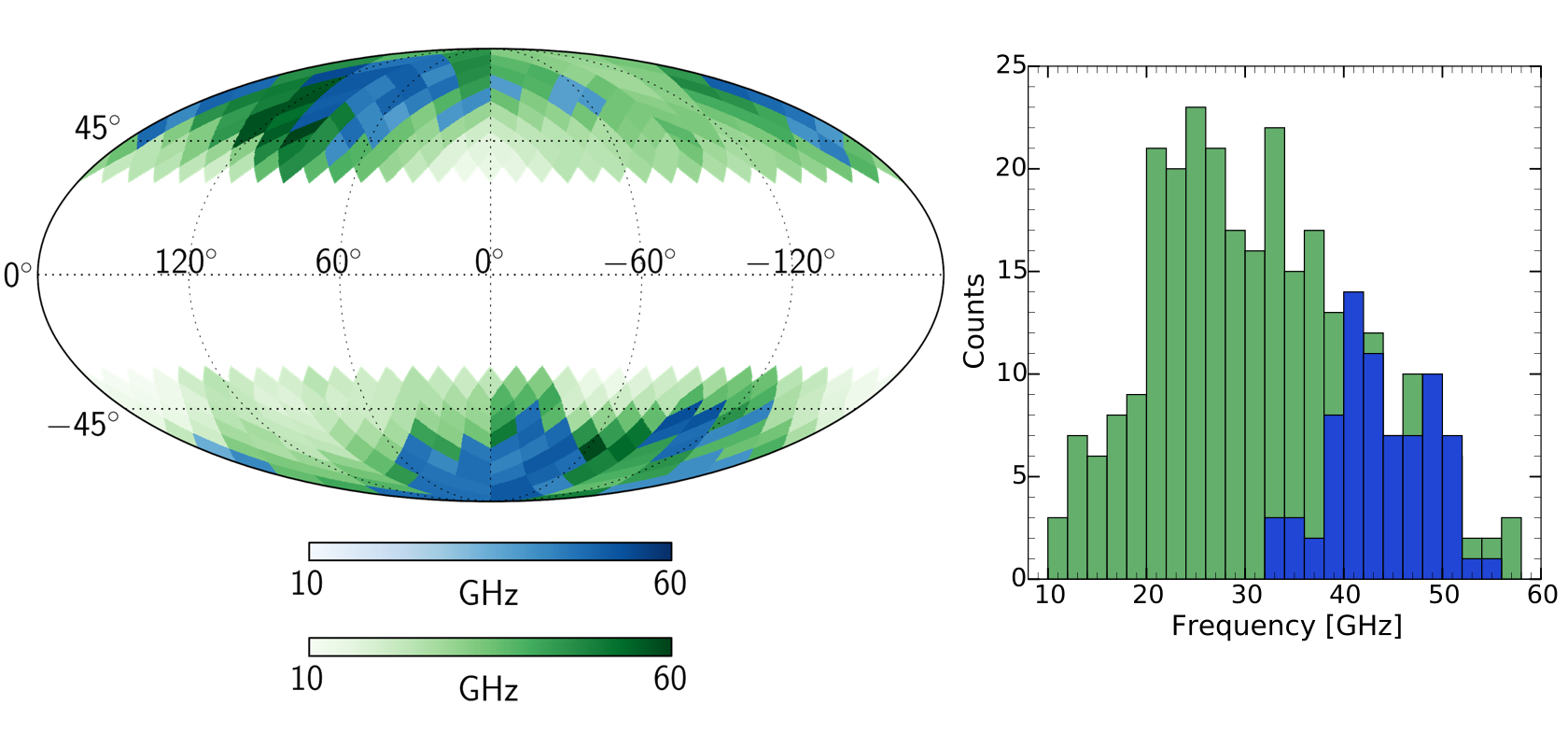}
      
            \caption{Frequency at which synchrotron and thermal dust contamination to CMB $B$-modes drops below $r=0.005$ (top and bottom panel respectively). Green pixels for detections, grey pixels for upper limits, blue pixels for lower limits. 
              }
         \label{map_onlyONE}
   \end{figure*}

\subsection{Single foreground contribution}
We can use the same methodology described above to infer the frequencies at which only one type of foreground contaminates significantly the CMB $B$-mode signal, while the second is below a certain threshold. This information is of particular 
interest, since it can be used to optimize instrument design, understanding how many channels are needed, and at which frequency, to be able to disentangle the CMB signal from the Galactic emissions. \par
We start to estimate the frequency range where the synchrotron emission is negligible. To do this, we extrapolate the $BB$ synchrotron map in figure (\ref{l80_maps}), using the power-law SED described in the previous section We then look, in each region, for the minimum frequency at which $r_s<0.005$ (with the convention adopted above, $r_s$ is the amplitude of extrapolated synchrotron emission in the $\ell_{80}$ bandpower divided by the amplitude of the CMB $B$-modes with $r=1$ in the same bin). Results are reported on the map and the histogram in the top panel of Figure (\ref{map_onlyONE}). We can infer the actual value of this frequency only in the 32 regions where we detect a synchrotron $B$-mode signal (green pixels on map). For all the other sky patches we  put upper limits. Considering both detections and upper limits the retrieved frequencies vary between $\sim110$ and $\sim160$ GHz across the sky (with errors between $\sim3$ and $\sim10$ GHz). From this analysis, we can therefore 
conclude that, given the current knowledge we have on polarized synchrotron radiation, it is not possible to exclude that CMB $B$-modes measurements at frequency below $\sim110$ GHz are contaminated by this kind of Galactic emission at the level of $r=0.005$, in any region of the sky. Considering a most stringent threshold on the amplitude of synchrotron, with $r_s<0.001$, this frequency increases at $\sim150$ GHz.  \par
Similarly we  rescale the $BB$ thermal dust map in figure  (\ref{l80_maps}) to find, for each region, the maximum frequency at which $r_{d}<0.005$. Results are shown in the bottom panel of Figure (\ref{l80_maps}), for the regions where we have a detection of the thermal dust signal (green pixels) and for those where we can only put lower limits on the value of this frequency (blue pixels). Considering all sky regions, the maximum value we find is $57\pm12$ GHz, dropping to about 40 GHz if we impose $r_d<0.001$.\par
Table \ref{r_min_info} reports, for the regions of interest, the retrieved minimum and maximum frequencies at which $r_s<0.005$ and $r_d<0.005$ respectively.

\section{Discussion and conclusions}
\label{Sec_conc}

\begin{figure*}
   \centering
   \includegraphics[width=13 cm]{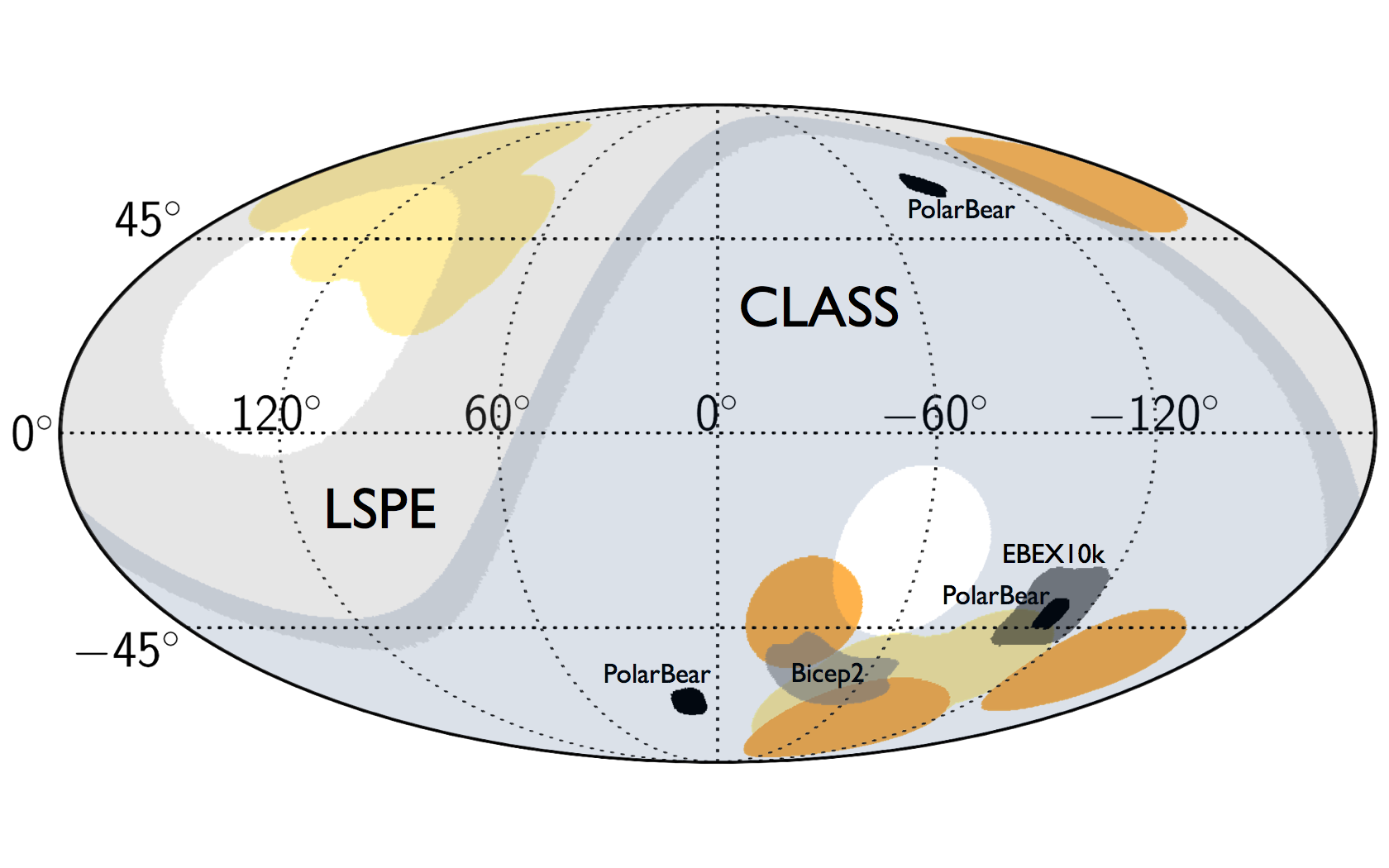}
            \caption{Observed or targeted sky patches for the quoted experiments (grey and black areas) versus locations of the regions described in Table \ref{r_min_info}. Dark yellow regions represent the first four entries of Table \ref{r_min_info}, light yellow color refers to the remaining sixteen.}
         \label{map_exp}
   \end{figure*}

We quantify the contamination from polarized diffuse Galactic synchrotron and thermal dust emissions to   
the $B$-modes of CMB anisotropies on the degree angular scale. Our analysis is based on the data made available from the 
\textit{Planck} and WMAP satellites, at Galactic latitudes with $|b|>20^{\circ}$, and exploit the {\tt Xpure} 
power spectrum estimator, which we validate on suitable simulations. 
We measure the synchrotron emission evaluating cross-spectra of the 30 GHz \textit{Planck}-LFI channel with the WMAP-$K$ band, centered at 22 GHz. The thermal dust emission is quantified using the 353 GHz channel of the HFI. We assign uncertainties to power spectra using the noise covariance matrices of the adopted maps. \par
We focus on the $\ell_{80}$ bandpower in the angular domain, corresponding to the angular scales where the CMB $B$-modes recombination bump peaks and where, currently, sub-orbitals experiments are focusing to constraint $r$. In particular we evaluate power spectra of foreground emissions in 352 small circular regions of the sky with $f_{\text{sky}}\simeq1.5\%$. In 28 sky regions, both synchrotron and 
dust spectra are measurable, and there, extrapolation in frequency suggests a 
minimum foreground level which is equivalent to a tensor-to-scalar ratio $r_{FG, \text{min}}$ between
$\sim0.06$ and $\sim1.0$, in the frequency interval between about 60 and 100 GHz. In all the regions where synchrotron or dust  $B$-modes polarized signal is not detectable at $3\sigma$ significance, we put upper limits on the minimum foreground contamination, finding values of $r_{FG, \text{min}}$ between $\sim0.05$ and $\sim1.5$, in the frequency range 60-90 GHz. \par
Table \ref{r_min_info} summarized the obtained results in twenty regions:

\begin{itemize}
\item the four regions where both thermal dust and synchrotron are detected in the $\ell_{80}$ bin and with recovered value  $r_{FG, \text{min}}<0.12$;
\item the sixteen regions where the upper limit on $r_{FG, \text{min}}$ is $\lesssim0.06$.
\end{itemize}
\par
We also estimate the frequency range where the only significant contamination arises from a single foreground component. 
On the high frequency side, where the dust emission dominates, the synchrotron is estimated to be negligible 
(with $r_s<0.005$) only at frequencies larger than 160 GHz (considering all the regions analyzed). The upper limits we put on this frequency are never less than $\sim110$ GHz. On the low frequency side, the dust drops at the same level in the 10-60 GHz frequency range. 

The same twenty sky patches reported in Table \ref{r_min_info} are also displayed on map in Figure \ref{map_exp}, together with several regions which have been, or will be, observed by current and future ground-based or ballon-borne CMB experiments. In particular we consider the cases of the LSPE \citep{LSPE12} 
and EBEX10K \citep{EBEX} proposed balloons, the CLASS ground-based experiment \citep{CLASS}, as well as the areas observed by BICEP2/Keck Array \citep{BICEP2, B2K} and PolarBear \citep{PolarBear14}. \par

In the Southern Galactic Hemisphere we find a large region of the sky characterized by 
low upper limits on $r_{FG, \mathrm{min}}$. This confirms that this area appears to be 
among the cleanest in the sky. However, in the proximity of this region (in particular at the edges of the
area observed by BICEP2) we have a few patches where synchrotron 
B-modes are detected with an extrapolated amplitude corresponding
to $r_s\approx 0.03$ at about 90 GHz. This is consistent with the
recent analysis of Keck-BICEP2 collaborations \citep{B2K2015} that improves 
the current upper limits on tensor-to-scalar ratio to $r<0.09$, highlighting a small 
synchrotron contribution in this field (the first panel in Figure 3
of \citealp{B2K2015}, in particular, shows a small excess in the cross-correlation between 
WMAP $K$-band data with Keck and BICEP2 measurements).
Moreover, close to this area of the sky, we detect synchrotron $E$-modes 
(see Figure \ref{l80_maps}), which can trace possible $B$-modes at a level below the sensitivity
of our analysis. In summary, our analysis confirms that this region in the Southern Galactic Hemisphere
contains little contribution from polarized synchrotron emissions compared
to other regions in the sky. The presence of these low residuals, however,
should not be neglected in view of experimental efforts aimed at detecting
$r$ at levels $r < 0.03$ at frequencies below $\sim100$ GHz.
Similarly, there is a large area with low upper limits in the North Galactic hemisphere, where, we do not detect synchrotron $B$-modes signal but we observe synchrotron $E$-modes.\par

The CLASS experiment and the low frequency instrument onboard the LSPE ballon \citep{STRIP12} will observe a large fraction of the sky ($\sim60\%$ and $\sim30\%$ of the sky respectively) with high sensitivity at low frequencies (around 40 GHz), partially covering also the two sky areas described above. These measurements will give therefore the possibility to improve the characterization of synchrotron polarized radiation.\par 
In conclusion, our results indicate that, with the current sensitivity at low frequency, it is not possible to exclude the presence of synchrotron contamination to CMB cosmological $B$-modes measurements at frequency $\lesssim100$ GHz anywhere. At these frequency more accurate data are essential in order to understand the synchrotron polarized component, and eventually remove its contamination to CMB measurements through foreground cleaning. Restricting the observations to higher frequencies ($\gtrsim110$ GHz) and focusing only on dust contamination would be an option to target a CMB $B$-mode signal with $r\simeq0.01$, since in several sky regions the synchrotron emission drops at $r_s<0.005$ at these frequency. Nevertheless, the observation of a more fainter GWs signal (with $r<0.01$)  will require instruments with both low and high frequency channels to monitor, with high accuracy, both kinds of foreground emission.

\begin{acknowledgements}
We gratefully acknowledge support from ASI/INAF Agreement 2014-024-R.0 for the
Planck LFI Activity of Phase E2.
NK acknowledges Maurizio Tomasi for careful reading the manuscript.
CB acknowledges partial support by the INDARK INFN Grant.

\end{acknowledgements}

%
%

\bibliographystyle{aa} 
\bibliography{FG_bib} 

\end{document}